\documentclass[journal]{IEEEtran}
\usepackage{cite}
\usepackage[pdftex]{graphicx}
\usepackage{amsmath}
\usepackage{algorithmic}
\usepackage{array}
\usepackage{multirow}
\usepackage{amssymb}
\usepackage{cases}

\usepackage{url}
\begin{document}

\title{MMV-Net: A Multiple Measurement Vector Network for Multi-frequency Electrical Impedance Tomography}

\author{Zhou~Chen,~\IEEEmembership{Student~Member,~IEEE}, Jinxi~Xiang,~\IEEEmembership{Student~Member,~IEEE,} Pierre Bagnaninchi,
        and~Yunjie~Yang,~\IEEEmembership{Member,~IEEE}
\thanks{Z. Chen and Y. Yang are with the Intelligent Sensing, Analysis and Control Group, Institute for Digital Communications, School of Engineering, The University of Edinburgh, Edinburgh, UK, EH9 3JL (E-mail: y.yang@ed.ac.uk).}
\thanks{J. Xiang is with the the Department of Precision Instrument, Tsinghua University, Beijing 100084, China.}
\thanks{P. Bagnaninchi is with the Centre for Regenerative Medicine, Institute for Regeneration and Repair, The University of Edinburgh, Edinburgh EH16 4UU. (E-mail: Pierre.Bagnaninchi@ed.ac.uk).}
\thanks{Manuscript received xx, 2021.}}


\maketitle
\begin{abstract}
Multi-frequency Electrical Impedance Tomography (mfEIT) is an emerging biomedical imaging modality to reveal frequency-dependent conductivity distributions in biomedical applications. Conventional model-based image reconstruction methods suffer from low spatial resolution, unconstrained frequency correlation and high computational cost. Deep learning has been extensively applied in solving the EIT inverse problem in biomedical and industrial process imaging. However, most existing learning-based approaches deal with the single-frequency setup, which is inefficient and ineffective when extended to address the multi-frequency setup. In this paper, we present a Multiple Measurement Vector (MMV) model based learning algorithm named MMV-Net to solve the mfEIT image reconstruction problem. MMV-Net takes into account the correlations between mfEIT images and unfolds the update steps of the Alternating Direction Method of Multipliers (ADMM) for the MMV problem. The non-linear shrinkage operator associated with the weighted $l_{2,1}$ regularization term is generalized with a cascade of a Spatial Self-Attention module and a Convolutional Long Short-Term Memory (ConvLSTM) module to capture intra- and inter-frequency dependencies. The proposed MMV-Net was validated on our \textit{Edinburgh mfEIT Dataset} and a series of comprehensive experiments. All reconstructed results show superior image quality, convergence performance and noise robustness against the state of the art.
\end{abstract}

\IEEEpeerreviewmaketitle

\section{Introduction}
\IEEEPARstart{B}{io-impedance} as an indicator of the physiological status of biological tissues varies with frequency. Electrical Impedance Tomography (EIT) is a non-intrusive and non-destructive imaging modality for revealing the cross-sectional conductivity distribution from a sequence of boundary current injections and induced differential voltage measurements \cite{eittissuemea:Adler}. The EIT inverse problem is to reconstruct the conductivity inside the Region Of Interest (ROI) based on the voltage measurements. EIT is fast, low-cost, portable, non-intrusive, label-free and radiation-free, making it an up-and-coming candidate in biomedical imaging. Emerging applications include functional lung imaging \cite{b3}, stroke diagnosis \cite{mfeitdata:Goren, mfeit:McDermott}, and biological tissue imaging \cite{tissueeit:Adler,fdeit3dhemisphere:Sujin}. As the impedance spectra of biological tissues is frequency-dependent, differences in electrical properties between various tissues can be exploited to benefit physiological and pathological diagnostics for tissue differentiation, early cancer detection, and tumor or stroke imaging \cite{tissueeit:Adler, freqspecp:Yang}. This motivates the development of multi-frequency EIT (mfEIT) \cite{mfeit23D:Yang}, which measures the bio-impedances under different frequencies of interest and employ them to reconstruct a set of multi-frequency conductivity images that are related to tissue properties.


The image reconstruction problem of mfEIT concerns the simultaneous reconstruction of multiple conductivity images of selected frequencies. The problem is fundamentally challenging on account of its non-linearity, severe ill-posedness, sensitivity to modelling and measurement errors and high computational cost. However, many existing literature focuses on advancing mono-frequency EIT image reconstruction algorithms based on the single measurement vector (SMV) model \cite{sparsity:Jin, ags:Yang, sbl:Liu, shapecons:Ren, shapeboolean:Liu}. The effort in developing effective and efficient mfEIT image reconstruction algorithms has been relatively limited. A straightforward way is to recover the image under each frequency individually by using the SMV-based methods while ignoring the correlations among conductivity images over the selected frequencies. In contrast, Multiple Measurement Vector (MMV) model \cite{mmv:Ziniel} as an extension of SMV considers the mfEIT measurements as a whole and can simultaneously reconstruct multiple conductivity images of given frequencies with higher quality by exploiting the inherent correlation between multi-frequency images. Specifically, it enforces the joint-sparsity constraint to each pixel within ROI to exploit the spatial structure with the hypothesis that solutions for all frequencies have common profiles but unnecessarily similar magnitudes. To optimize the MMV model, the MMV-ADMM algorithm adopts the extensively utilized Alternating Direction Method of Multipliers (ADMM) \cite{ADMM1, ADMM2} and has been introduced in hyperspectral imaging \cite{abundancees:Qu} and multi-frequency Electrical Capacitance Tomography (ECT) \cite{admmmmv:Zhang}. Alternatively, Liu \textit{et al.} \cite{mtsasbl:Liu} and Xiang \textit{et al.} \cite{mfemt:Xiang} extended the SMV-based Sparse Bayesian Learning (SBL) framework based on the MMV model for multi-frequency tomographic imaging. However, the MMV-SBL approaches rely on heavily the sparsity assumption, leading to considerably degraded performance in non-sparse scenarios. In addition, the computational cost of MMV-ADMM and MMV-SBL is considerable, preventing their wide adoption in biomedical applications that desire real-time imaging capability.

Recently, deep learning has proven its effectiveness for medical tomographic image reconstruction with significant image quality improvement \cite{imagenf:Wang}. Composed of a stack of layers, deep neural networks learn complicated functions automatically from large-scale training datasets without requiring manually designed priors. Learning-based methods are also desirable for real-time imaging due to the faster execution time against conventional model-based algorithms. Existing learning-based EIT image reconstruction algorithms can be classified into three categories \cite{FISTANet:Xiang}: (a) fully learning approaches that directly map measurement data to a conductivity image \cite{cnnert:Tan}; (b) image post-processing approaches that employ a trained network to eliminate artifacts of a preliminary conductivity image obtained from model based algorithms \cite{bar:Hamilton,Dominant:Wei}; (c) model-based deep learning approaches that unroll a finite number of iterations of the model based methods into a network, e.g. FISTA-Net \cite{FISTANet:Xiang}, MoDL \cite{Modl:Aggarwal}, ADMM-CSNet \cite{Admm-csnet:Yang} and ISTA-Net \cite{Ista-net:Zhang}. Generic deep networks in a) and b) via end-to-end training lack interpretability and the underlying structure neglects the physical processes of the problem. In contrast, the unrolling approaches in c) are capable of incorporating the advantages of physical model based methods and deep neural networks. 

Despite the advancement, learning-based methods for mfEIT image reconstruction remains an open problem. To address the research gap this paper proposes a model-based deep learning approach for high-performance mfEIT image reconstruction. The proposed approach, named as MMV-Net, unrolls the MMV-ADMM algorithm into a single pipeline (see Fig.~\ref{fig:network}). MMV-Net is composed of multiple blocks, each of which corresponds to one iteration. The non-linear shrinkage operation in each block is approximated and generalized by a deep network. MMV-Net is fundamentally different from ISTA-Net \cite{ista:Bioucas} and FISTA-Net \cite{fista:Beck}. Although MoDL, ADMM-CSNet and MMV-Net have similar structure based on ADMM, MoDL and ADMM-CSNet are exclusively designed for SMV-based imaging whereas MMV-Net goes beyond to provide simultaneously multiple conductivity images constrained by frequency correlations to promote image quality. The main contributions of this paper are as follows: 
\begin{enumerate}
\item A novel model-based deep learning approach is proposed for simultaneous mfEIT image reconstruction. The proposed MMV-Net tackles the inherent limitations of the conventional MMV-ADMM approach. Parameters across all iteration blocks are shared and learned through end-to-end training.
  
\item A dedicated network is developed to substitute the non-linear shrinkage operator, which learns a more general regularizer to incorporate the in essence spatial and frequency correlations between mfEIT images. The unique design could boost the reconstruction performance of mfEIT.

\item The proposed MMV-Net has much fewer parameters than the state-of-the-art model-based learning approches, e.g. MoDL \cite{Modl:Aggarwal} and FISTA-Net \cite{FISTANet:Xiang} (by 12.8 and 8.5 times respectively), making it much easier to train even with limited dataset.

\item  \textit{Edinburgh mfEIT Dataset} is generated for mfEIT image reconstruction. The dataset mimics tissue engineering applications and comprises $4 \times 12,414$ randomly generated multi-object, multi-conductivity phantoms at four frequencies.

\item MMV-Net is thoroughly evaluated on \textit{Edinburgh mfEIT Dataset} and various real-world experiments, which significantly outperforms the state of the art, including MMV-ADMM \cite{admmmmv:Zhang}, MoDL \cite{Modl:Aggarwal}, and FISTA-Net \cite{FISTANet:Xiang}.

\end{enumerate}

The remainder of this paper is organized as follows. In Section \ref{sec:Methodology}, we present the problem formulation of mfEIT and elaborate the proposed MMV-Net. Section \ref{sec:EandR} describes the \textit{Edinburgh mfEIT Dataset}, experiment implementation, and simulation and experimental results. Section \ref{sec:Conclusion} draws conclusions and discusses future work.

\section{Methodology}\label{sec:Methodology}

\subsection{Multi-frequency EIT} \label{mmvfoundation}
Consider difference imaging of mfEIT \cite{nonlineardiff:Liu}, voltage changes $\Delta \textbf{V} \in \mathbb{R}^{m \times l}$ at a set of excitation frequencies $ \{f_1, f_2, \dots, f_l\}$ are measured to reconstruct the conductivity changes $\Delta \pmb{\sigma} \in \mathbb{R}^{n\times l}$. The Multiple Measurement Vector (MMV) model \cite{mmv:Ziniel} of mfEIT linearly approximates the relationship between $\Delta \textbf{V}$ and $\Delta \pmb{\sigma}$ by
\begin{equation} \label{eq:mmv}
\Delta \textbf{V} = \textbf{A} \Delta \pmb{\sigma}
\end{equation}
where $\textbf{A} \in \mathbb{R}^{m \times n}$ ($m\ll n$) denotes the sensitivity matrix.

We define $\Delta \textbf{V}=[\Delta \textbf{v}_{f_1}, \Delta \textbf{v}_{f_2}, \dots, \Delta \textbf{v}_{f_l}]$, where $\Delta \textbf{v}_{f_i} \in \mathbb{R}^{m \times 1} (i=1, \dots,l)$ denotes the $i^{th}$ column of $\Delta \textbf{V}$. The first method to obtain voltage changes leverages Time-Difference (TD) measurements \cite{eitreview:Brown}, i.e. $ \Delta \textbf{v}_{f_i}= \textbf{v}_{f_i}(t_1) - \textbf{v}_{f_i}(t_0) $, which requires mfEIT measurements at two time instants $t_0$ and $t_1$, i.e. $\textbf{v}_{f_i}(t_1), \textbf{v}_{f_i}(t_0) \in \mathbb{R}^{m \times 1}$. Another prevailing method is to utilize Frequency-Difference (FD) measurements \cite{detectinclu:Harrach}. The FD approach employs voltage changes at different frequencies, i.e. $ \Delta \textbf{v}_{f_i}= \textbf{v}_{f_i} - \textbf{v}_{f_0}$, where $f_0$ is the reference frequency.

For convenience, we use $ \textbf{B}$ and $ \textbf{X}$ as substitutes for $\Delta \textbf{V}$ and $ \Delta \pmb{\sigma}$ respectively in the rest of the paper. Typically, the MMV model based mfEIT image reconstruction problem can be solved by addressing the constrained optimization problem:

\begin{equation} \label{eq:optgeneral}
\begin{cases}
     \displaystyle\min_{ \textbf{X}} \mathcal{R}( \textbf{X}) \\
    s.t.\,\textbf{A} \textbf{X} = \textbf{B}
  \end{cases}
\end{equation}
where $\mathcal{R}(\cdot)$ denotes the regularization function, which encodes the prior knowledge of the conductivity distribution $ \textbf{X}$. The MMV model uses the weighted $l_{2,1}$ regularization to facilitate the joint sparsity \cite{admmmmv:Zhang,groupsparseopti:Deng}:
\begin{equation} \label{eq:opt}
\begin{cases}
     \displaystyle\min_{ \textbf{X}} \| \textbf{X} \|_{w,2,1} := \sum_{i=1}^{n} w_i\left\lVert \textbf{X}_{i} \right\rVert _2 \\
    s.t.\,\textbf{A} \textbf{X} = \textbf{B}
  \end{cases}
\end{equation}
where $w_i$($i=1, \dots, n$) is a positive scalar and $\textbf{X}_{i}$ is the $i^{th}$ row of $ \textbf{X}$.

\subsection{MMV-ADMM for mfEIT image reconstruction}
The MMV-based mfEIT-image-reconstruction problem in \eqref{eq:opt} can be efficiently solved by using the classic Alternating Direction Method of Multipliers (ADMM) {\cite{ADMM1, ADMM2}}. By introducing an auxiliary vector $\textbf{Z} \in \mathbb{R}^{n\times l}$, the problem in \eqref{eq:opt} is equivalent to
\begin{equation}\label{eq:auxi}
\begin{cases}
     \displaystyle\min_{\textbf{X}, \textbf{Z}}  \| \textbf{Z} \|_{w,2,1} \\
    s.t.\, \textbf{Z}=\textbf{X}, \textbf{A} \textbf{X} = \textbf{B}.
  \end{cases}
\end{equation}

The augmented Lagrangian problem of \eqref{eq:auxi} is
\begin{equation}\label{eq:Lag}
\begin{aligned}
\displaystyle\min_{\textbf{X}, \textbf{Z}}\,  \| \textbf{Z} \|_{w,2,1}
-\pmb{\Lambda}_{1}^{\mathrm{T}}(\textbf{Z}& -\textbf{X})
+\frac{\beta_{1}}{2}\|\textbf{Z}-\textbf{X}\|_{2}^{2}-\\
&\pmb{\Lambda}_{2}^{\mathrm{T}}(\textbf{A} \textbf{X}-\textbf{B})
+\frac{\beta_{2}}{2}\|\textbf{A} \textbf{X}-\textbf{B}\|_{2}^{2}
\end{aligned}
\end{equation}
where $\pmb{\Lambda}=\{\pmb{\Lambda}_{1} \in \mathbb{R}^{n\times l},  \pmb{\Lambda}_{2} \in \mathbb{R}^{m\times l} \}$ are Lagrangian multipliers and ${\beta}=\{{\beta}_{1},{\beta}_{2} >0\} $ are penalty parameters. The ADMM is then applied to solve \eqref{eq:Lag} through the following steps:
\begin{equation}\label{eq:optsub}
\begin{cases}
\underset{\textbf{X}}{\arg\min}\,\pmb{\Lambda}_{1}^{\mathrm{T}}\textbf{X}+\frac{\beta_{1}}{2}\|\textbf{Z}-\textbf{X}\|_{2}^{2}- \pmb{\Lambda}_{2}^{\mathrm{T}}\textbf{A}\textbf{X}\\
\qquad\qquad +\frac{\beta_{2}}{2}\|\textbf{A} \textbf{X}-\textbf{B}\|_{2}^{2}, \\

\underset{\textbf{Z}}{\arg \min}\, \| \textbf{Z} \|_{w,2,1}-\pmb{\Lambda}_{1}^{\mathrm{T}}\textbf{Z}+\frac{\beta_{1}}{2}\|\textbf{Z}-\textbf{X}\|_{2}^{2}, \\

\pmb{\Lambda}_{1}\xleftarrow{}\pmb{\Lambda}_{1}-{\gamma}_1{\beta}_{1}(\textbf{Z}-\textbf{X}),\\
\pmb{\Lambda}_{2}\xleftarrow{}\pmb{\Lambda}_{2}-{\gamma}_2{\beta}_{2}(\textbf{A} \textbf{X}-\textbf{B}),
\end{cases}
\end{equation}
where ${\gamma}=\{{\gamma}_{1},{\gamma}_{2} >0\} $ are step lengths for the Lagrangian multipliers.

The first step in \eqref{eq:optsub} is a convex quadratic problem, which has a closed-form solution:
\begin{equation}\label{eq:cfsolution}
\begin{aligned}
\textbf{X}=({\beta}_{1}\textbf{I}+{\beta}_{2}\textbf{A}^{\mathrm{T}}\textbf{A} )^{-1}&({\beta}_{1}\textbf{Z}-\\
&\pmb{\Lambda}_{1}+{\beta}_{2}\textbf{A}^{\mathrm{T}}\textbf{B}+\textbf{A}^{\mathrm{T}} \pmb{\Lambda}_{2})
\end{aligned}
\end{equation}
where $\textbf{I} \in \mathbb{R}^{n \times n}$ is an identity matrix.

To avoid large matrix inversion and reduce the computation cost, we in this work adopt the gradient descent method as a substitute for \eqref{eq:cfsolution}:
\begin{equation} \label{eq:gdsolution}
\textbf{X} \xleftarrow{} \textbf{X} -\eta {\nabla}_{\textbf{X};\textbf{Z}, \pmb{\Lambda}_{1}, \pmb{\Lambda}_{2}}
\end{equation}
where $\eta$ is the step size, and ${\nabla}_{\textbf{X};\textbf{Z}, \pmb{\Lambda}_{1}, \pmb{\Lambda}_{2}}$ is the gradient of the first step in \eqref{eq:optsub} with respect to $\textbf{X}$ given $\{ \textbf{Z}, \pmb{\Lambda} \}$, which is defined by:
\begin{equation}\label{eq:gradi}
\begin{aligned}
{\nabla}_{\textbf{X};\textbf{Z}, \pmb{\Lambda}_{1}, \pmb{\Lambda}_{2}}=({\beta}_{1}\textbf{I}+&{\beta}_{2}\textbf{A}^{\mathrm{T}}\textbf{A} )\textbf{X}  - ({\beta}_{1}\textbf{Z}-\\
&\pmb{\Lambda}_{1}+{\beta}_{2}\textbf{A}^{\mathrm{T}}\textbf{B}+\textbf{A}^{\mathrm{T}} \pmb{\Lambda}_{2})
\end{aligned}
\end{equation}

Thus, to solve (4) with ADMM, we have the following updates at the $k^{th}$ iteration:
\begin{subnumcases}{}
&$\textbf{X}^{(k)}=\textbf{X}^{(k-1)} -\eta {\nabla}_{\textbf{X}^{(k-1)};\textbf{Z}^{(k-1)}, \pmb{\Lambda}_{1}^{(k-1)} , \pmb{\Lambda}_{2}^{(k-1)} },$\label{eq:solusub1}\\
&$\textbf{Z}^{(k)}=\mathcal{S}(\textbf{X}^{(k)}+\frac{1}{\beta_{1}}\pmb{\Lambda}_{1}^{(k-1)},\frac{w}{\beta_{1}}),$\label{eq:solusub2}\\
&$\pmb{\Lambda}_{1}^{(k)}=\pmb{\Lambda}_{1}^{(k-1)}+{\gamma}_1{\beta}_{1}(\textbf{X}^{(k)}-\textbf{Z}^{(k)}),$\label{eq:solusub3}\\
&$\pmb{\Lambda}_{2}^{(k)}=\pmb{\Lambda}_{2}^{(k-1)}+{\gamma}_2{\beta}_{2}(\textbf{B}-\textbf{A} \textbf{X}^{(k)}),$\label{eq:solusub4}
\end{subnumcases}
where $k\in \{1\dots K_s\}$ is the iteration index, and $\mathcal{S}(\cdot)$ represents a row-wise shrinkage operator associated with the weighted $l_{2,1}$ regularization:
\begin{equation}\label{eq:l21}
\begin{aligned}
\textbf{Z}_i^{(k)}&=\max \left\{\left\|\textbf{X}_i^{(k)}+\frac{1}{\beta_{1}}(\pmb{\Lambda}_{1})_i^{(k-1)}\right\|_{2}-\frac{w_i}{\beta_{1}}, 0\right\}\cdot\\
&\frac{\textbf{X}_i^{(k)}+\frac{1}{\beta_{1}}(\pmb{\Lambda}_{1})_i^{(k-1)}}{\left\|\textbf{X}_i^{(k)}+\frac{1}{\beta_{1}}(\pmb{\Lambda}_{1})_i^{(k-1)}\right\|_{2}}, for\ i=1, \dots, n.
\end{aligned}
\end{equation}

\begin{figure}[!t]
\centerline{\includegraphics[width=\linewidth]{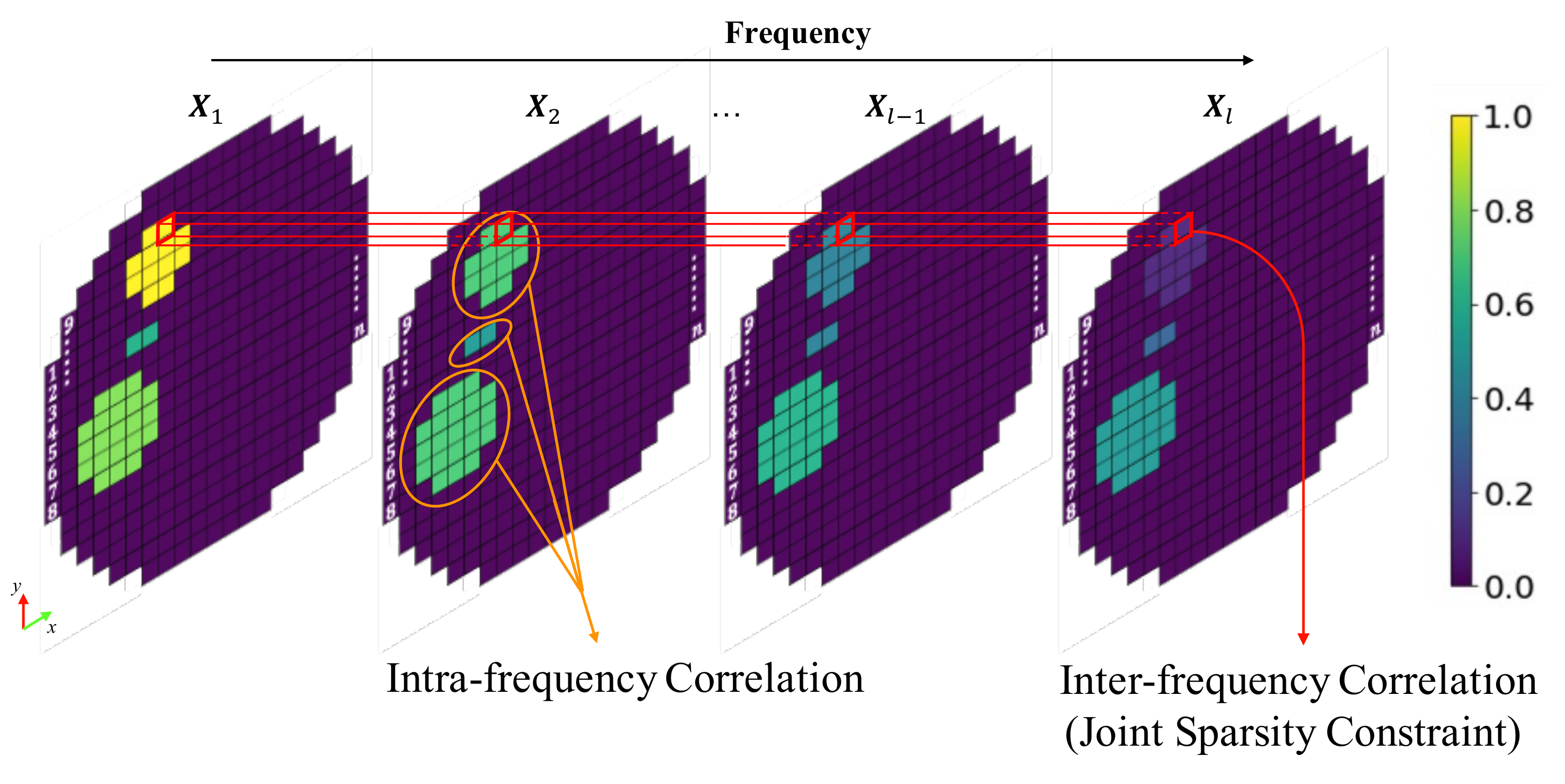}}
\caption{Illustration of intra- and inter-frequency correlation between mfEIT images.}
\label{fig:correlations}
\end{figure}

In this work, we propose to approximate $\mathcal{S}(\cdot)$ through the data-driven method. Instead of directly applying \eqref{eq:l21} to learn the joint sparsity only, MMV-Net aims to learn a more general $\mathcal{S}(\cdot)$ by taking into account both the spatial correlation and the inherent correlation across different frequencies (see illustration in Fig.~\ref{fig:correlations}), to simultaneously reconstruct the multi-frequency conductivity distributions effectively and efficiently.


\subsection{MMV-Net for mfEIT image reconstruction}

\begin{figure*}[!t]
\centerline{\includegraphics[width=\linewidth]{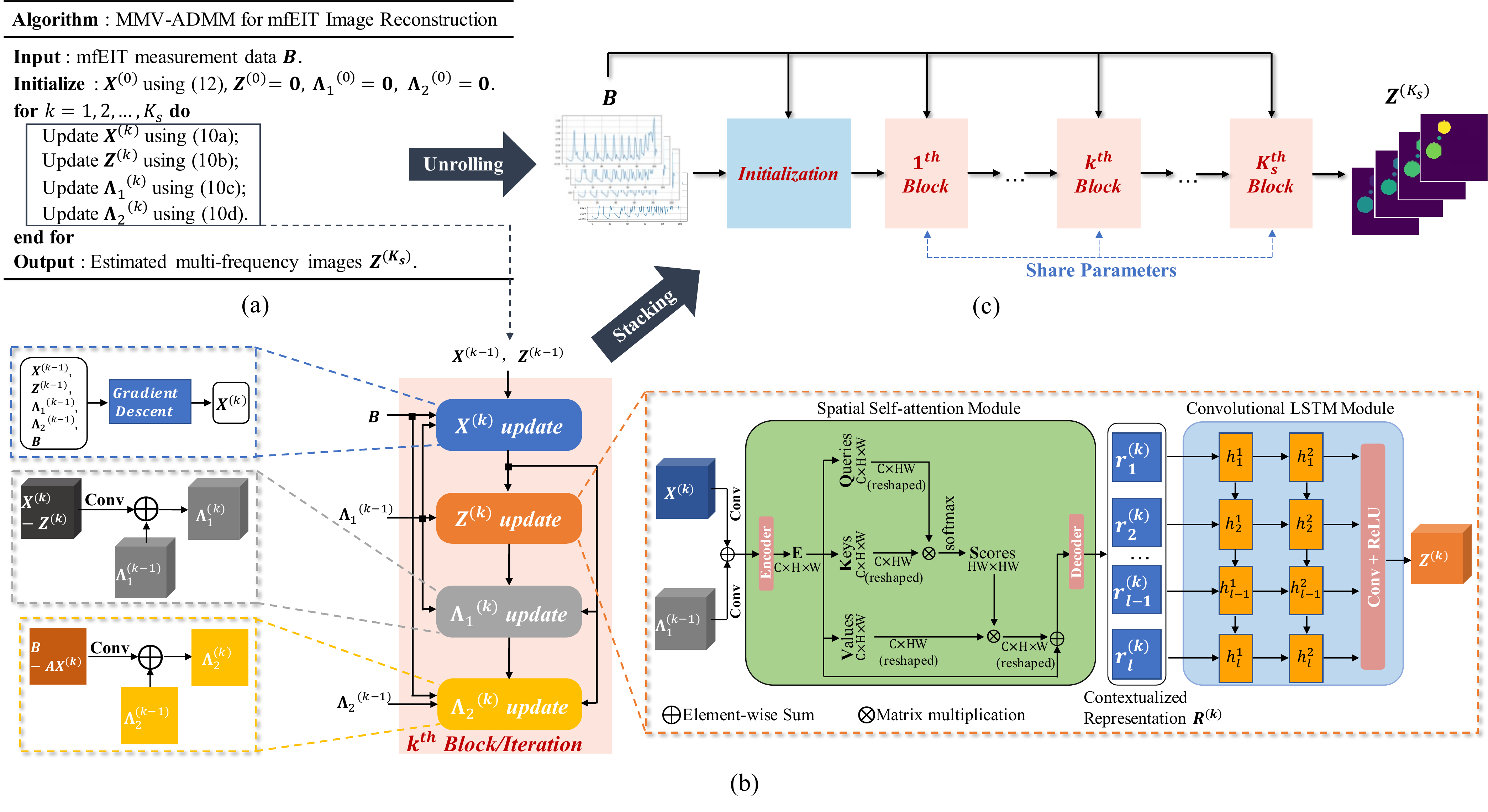}}
\caption{(a) The iterative MMV-ADMM algorithm. (b) Illustration of the four updating steps at the $k^{th}$ iteration corresponding to \eqref{eq:solusub1}-\eqref{eq:solusub4}. (c) Overall architecture of the proposed MMV-Net. MMV-Net is an unrolled architecture for $K_s$ iterations. It alternates among the four update steps. These update steps share parameters across all iterations. 
}
\label{fig:network}
\end{figure*}

For the MMV-based mfEIT image reconstruction problem, the conventional ADMM (MMV-ADMM) has limitations in three respects: a) typically it repeats hundreds of iterations to achieve the optimum, which degrades the computational efficiency to a great extent \cite{abundancees:Qu,admmmmv:Zhang}; b) the non-linear shrinkage operator $\mathcal{S}(\cdot)$ is only valid for specific image patterns (e.g. sparsity \cite{sparsemmv:Cotter}, group sparsity \cite{admmmmv:Zhang}); c) it is non-trivial to fine-tune the algorithm parameters $\{\beta, \gamma, \eta \} $.

To address the above issues, we propose a deep architecture named MMV-Net for MMV-based mfEIT image reconstruction by unrolling the iterative MMV-ADMM algorithm. MMV-Net combines the model-based method and the deep neural network for mfEIT image reconstruction to exploit advantages from both sides. We map the four update procedures in \eqref{eq:solusub1}-\eqref{eq:solusub4} into an unfolded data flow with ${{K_s}} $ iterations as illustrated in Fig.~\ref{fig:network}.

The input of the MMV-Net is the mfEIT measurement data $ \textbf{B}$. $\textbf{Z}^{(0)}$, $\pmb{\Lambda}_{1}^{(0)}$, and $\pmb{\Lambda}_{2}^{(0)}$ are initialized as zeros. ${\textbf{X}^{(0)}}$ is obtained by employing the one-step Gaussian Newton solver with the Laplacian filter \cite{GREIT:Adler}:
\begin{equation}\label{eq:x0}
\textbf{X}^{(0)}=(\textbf{A}^{\mathrm{T}}\textbf{A}+\lambda \textbf{L}^{\mathrm{T}}\textbf{L})^{-1}\textbf{A}^{\mathrm{T}}\textbf{B}
\end{equation}
where ${\textbf{L}}$ is the Laplacian matrix, and $\lambda$ is the regularization factor. The initialization results are then utilized to generate the final multi-frequency conductivity images $\textbf{Z}^{({K_s})} $ after ${{K_s}} $ iterations. The subsequent part of MMV-Net comprises of ${{K_s}}$ blocks, where the $k^{th}$ block corresponds to the $k^{th}$ iteration of the MMV-ADMM algorithm. Each block consists of four update steps corresponding to one iteration of MMV-ADMM in \eqref{eq:solusub1}-\eqref{eq:solusub4}, including gradient descent ($\textbf{X}$) update, auxiliary variable ($\textbf{Z}$) update, the first multiplier ($\pmb{\Lambda}_{1}$) update, and the second multiplier ($\pmb{\Lambda}_{2}$) update. The following parts discuss the four update steps at the $k^{th}$ iteration in detail.

\subsubsection{$\textbf{X}^{(k)}$ \textbf{update}}
This update step implements the gradient descent method. It generates immediate results $\textbf{X}^{(k)}$. Given $\textbf{X}^{(k-1)}$, $\textbf{Z}^{(k-1)}$, $ \pmb{\Lambda}_{1}^{(k-1)}$, and $\pmb{\Lambda}_{2}^{(k-1)}$,  which are obtained from the previous $(k-1)^{th}$ iteration, the output $\textbf{X}^{(k)}$ is computed according to \eqref{eq:solusub1}.

\subsubsection{$\textbf{Z}^{(k)}$ \textbf{update}}

This step updates the auxiliary variable by unrolling the generalized non-linear operator $\mathcal{S}(\cdot)$ in \eqref{eq:solusub2}, where the prior knowledge is integrated. For mfEIT, it is crucial to learn intra- and inter-image correlations simultaneously with respect to frequency channels (see Fig.~\ref{fig:correlations}). With this purpose, we propose to design $\mathcal{S}(\cdot)$ as a deep neural network, in particular, a cascade of a Spatial Self-Attention (SSA) module and a Convolutional Long Short-Term Memory (ConvLSTM) module. Fig.~\ref{fig:network}(b) illustrates how \eqref{eq:solusub2} is mapped into a network. First, a more general combination of $\textbf{X}^{(k)}$ and $\pmb{\Lambda}_{1}^{(k-1)}$ is learned by two $1\times1\times1 $ convolutional layers respectively and an element-wise sum. Afterwards, we apply the SSA and ConvLSTM to improve reconstruction performance of $\textbf{Z}^{(k)}$. Let $Conv_1(\cdot)$ and $Conv_2(\cdot)$ be the two convolutional layers, $\mathcal{F}_{SSA}(\cdot)$ and $\mathcal{F}_{LSTM}(\cdot)$ be the function of SSA and ConvLSTM, respectively, then \eqref{eq:solusub2} can be reformulated as:
\begin{equation}\label{eq:SSA}
\textbf{Z}^{(k)}=\mathcal{F}_{LSTM}(\mathcal{F}_{SSA}(Conv_1(\textbf{X}^{(k)})+Conv_2(\pmb{\Lambda}_{1}^{(k-1)})))
\end{equation}

The detailed design of the SSA and ConvLSTM is elaborated as follows.

\textbf{Spatial self-attention module $\mathcal{F}_{SSA}(\cdot)$:} our previous work \cite{dlgs:Chen} observed that by explicitly learning the structural information of the conductivity image, significant image quality improvement can be achieved in terms of spatial resolution and accuracy. This idea is inherited by using a spatial self-attention module to determine the structural information under each frequency channel to extract inter-frequency correlations.

The family of attention modules is capable of modeling long-range dependencies in natural language processing and computer vision. As a variation of attention,  self-attention mechanism was firstly proposed to extract global dependencies of inputs for machine translation \cite{Attention:Vaswani}. \cite{Nonlocal:Wang} and \cite{Dualattention:Fu} extended the self-attention mechanism to video classification and image segmentation, respectively. Self-attention is usually inserted in a network and generates importance maps to refine the high-level feature maps. As a result, important regions can be focused on and feature representations are enriched with contextual relationships for intra-image compactness.

Inspired by the self-attention mechanism, the proposed SSA adopts the structure of a symmetric encoder–decoder network, embedded with a self-attention mechanism to the encoder output. Consider the fact that the reconstructed mfEIT images under all frequencies share the same structure, the encoder part first introduces a $3\times3 $ convolutional layer producing one output channel. Then, we apply two $2\times2$ convolutional layers with a stride of 2. Each layer is followed by a Batch Normalization layer and an ELU layer. The encoder outputs a feature representation $\textbf{E}$ with size of $C\times H\times W$, which is fed into three $3\times3 $ convolutional layers to generate feature maps queries $\textbf{Q}$, keys $\textbf{K}$ and values $\textbf{V}$, respectively. $\textbf{Q}$, $\textbf{K}$ and $\textbf{V}$ now have the size as $\textbf{E}$. They are all reshaped to $C\times P$, where $P = H\times W$. $\textbf{Q}$ and $\textbf{K}$ are multiplied and fed into a softmax layer to generate a score/attention map $\textbf{S}$ with size of $P\times P$. Afterwards $\textbf{S}$ and $\textbf{V}$ are multiplied and reshaped back to $C\times H\times W$. We then perform residual learning through a skip connection to $\textbf{E}$. Finally, the decoder is applied and it comprises of two $2\times2$ deconvolutional layers with a stride of 2, each of which is followed by a BatchNorm layer and a ELU layer. The final output is the contextualized representation of structure information $\textbf{R}^{(k)} \in \mathbb{R}^{m \times l}$.

\textbf{Convolutional LSTM module $\mathcal{F}_{LSTM}(\cdot)$:} based on the structure information $\textbf{R}^{(k)}$, we then attempt to learn the inter-frequency correlations by reconstructing the trend of the varying conductivity contrast along the frequency domain, meanwhile preserving the general structures learned from the SSA. 

We view the contextualized representation $\textbf{R}^{(k)}$ as a set of sequential images, i.e. $\{\textbf{r}_{i}^{(k)} \}_{i=1}^l$, where $\textbf{r}_{i}^{(k)} \in \mathbb{R}^{m}$ represents the $i^{th}$ column of $\textbf{R}^{(k)}$. To tackle this sequence-to-sequence (seq2seq) problem, Recurrent Neural Network (RNN) and LSTM models \cite{rnned:Cho,lstm:Donahue} are in a dominant position in the field of deep learning. One drawback of RNNs/LSTMs is that they require considerable memory to store intermediate cell gate parameters, especially for long sequences and high dimensional inputs, on account of the usage of full connections. Though powerful enough to capture temporal correlations, the fully-connected layers raise redundancy and distortion for spatial data. In contrast, Convolutional LSTM (ConvLSTM) \cite{ConvolutionalLSTM:Shi} is more computationally efficient as it replaces the fully-connected layers with convolutional layers. This operation further preserves spatial correlations with much less parameters and better generalization, meaning that we could employ more parameters to construct the SSA.

To learn the changes of conductivity contrast along the frequency, we take advantage of ConvLSTMs. The proposed ConvLSTM module has a stack of multiple ConvLSTM layers with a kernel size of $3\times3$. We set the layer number as two by default. We finally apply an additional $1\times1$ convolutional layer and a ReLU layer to generate $\textbf{Z}^{(k)}$.

\subsubsection{$\pmb{\Lambda}_{1}^{(k)}$ \textbf{update}} \label{sec:Lambda1Update}
The multiplier update step corresponds to \eqref{eq:solusub3}. As shown in Fig.~\ref{fig:network}(b), the residual $(\textbf{X}^{(k)}-\textbf{Z}^{(k)})$ first goes through a $1\times1\times1 $ convolutional layer, which is expected to learn ${\gamma}_1{\beta}_{1}$. Then we perform an element-wise sum operation between the residual and $\textbf{Z}^{(k-1)}$ to obtain the output $\pmb{\Lambda}_{1}^{(k)}$.

\subsubsection{$\pmb{\Lambda}_{2}^{(k)}$ \textbf{update}}
Fig.~\ref{fig:network}(b) also illustrates the multiplier update according to \eqref{eq:solusub4} with inputs of $\pmb{\Lambda}_{2}^{(k-1)}$ and $(\textbf{B}-\textbf{A} \textbf{X}^{(k)})$. Similar to the update of $\pmb{\Lambda}_{1}$, we decompose this operation to a $1\times1 $ convolutional layer to learn the product ${\gamma}_{2}{\beta}_{2}$ and an element-wise sum to generate $\pmb{\Lambda}_{2}^{(k)}$.

Given a training dataset $\mathcal{D}$ with $N_{\mathcal{D}}$ pairs of samples, we define the objective function as the Mean Square Error (MSE) between the predicted images $\textbf{Z}^{(K_s)}$ and the ground truth $\textbf{X}^{(gt)}$:
\begin{equation} \label{objfunc}
    \mathcal{L} = \frac{1}{N_{\mathcal{D}}}\sum_{( \textbf{B}, \textbf{X}^{(gt)}) \in \mathcal{D}}^{}
    \left\lVert \textbf{Z}^{(K_s)}- \textbf{X}^{(gt)} \right\rVert ^2.
\end{equation}

\subsection{Network Training}
We train the MMV-Net using PyTorch and employ Adam \cite{Adam:Kingma} for optimization with the batch size of 6. Similar to \cite{FISTANet:Xiang,Modl:Aggarwal,Admm-csnet:Yang,Ista-net:Zhang}, the non-linear operator and parameters $\pmb\Theta = \{Conv_1(\cdot), Conv_2(\cdot), \mathcal{F}_{SSA}(\cdot), \mathcal{F}_{LSTM}(\cdot), \beta, \gamma, \eta \} $ are all learned from training data, rather than hand tuning. We employ the parameter-sharing strategy to penalize the recursive network size for effective learning, where $\pmb\Theta $ of the MMV-Net are shared across all iterations. Inspired by the training approach in \cite{Modl:Aggarwal}, we adopt a three-step approach for training. We first train the auxiliary variable update ($\textbf{Z}$) to learn $\pmb\Theta_{\textbf{Z}} = \{Conv_1(\cdot), Conv_2(\cdot), \mathcal{F}_{SSA}(\cdot), \mathcal{F}_{LSTM}(\cdot)\} $. Then we train the entire parameters $\pmb\Theta$ for only one iteration, initialized with the previously learned parameters $\pmb\Theta_{\textbf{Z}}$. The trained parameters $\pmb\Theta$ with single iteration serve as a starting point of training the MMV-Net with multiple iterations.


\section{Experiments and Results}\label{sec:EandR}

\subsection{The Edinburgh mfEIT Dataset}

We established the \textit{Edinburgh mfEIT Dataset} (the dataset and code will be available in www.research.ed.ac.uk/en/datasets/) to train the proposed MMV-Net. It contains multiple imaging objects with continuously varying conductivity values along four frequencies ($l=4$) within a circular 16-electrode EIT sensor. The forward problem was solved by using COMSOL Multiphysics and Matlab. We adopt the adjacent measurement strategy \cite{collection:Brown} and a completed non-redundant measurement cycle contains $m=104$ voltage measurements. In solving the inverse problem, we divide the circular sensing region by a $64 \times 64$ quadrate mesh, which contains $n=3228$ pixels.

\begin{table}[!t]
\caption{Groups of simulated conductivity values at different frequencies.}
\begin{center}
\begin{tabular}{p{0.05\textwidth}  >{\centering}p{0.05\textwidth} >{\centering}p{0.05\textwidth} >{\centering}p{0.05\textwidth} >{\centering\arraybackslash}p{0.05\textwidth}
}
\hline
 Group Index & $f_1$ (S/m) &$f_2$ (S/m) &$f_3$ (S/m) & $f_4$ (S/m)  \\
\hline
\hline
{1} &0.01	&0.6&	1.2&	1.8\\

{2} & 0.4&	0.6&	0.8&	1.0\\

{3}  &0.8&	1.0&	1.2&	1.4\\
\hline

\end{tabular}
\label{table:condGroup}
\end{center}
\end{table}

The background substance is saline with a constant conductivity of 2 $S/m$, which does not change with frequency. One to three circular objects are simulated with their diameters randomly determined by the uniform distribution [0.05d, 0.3d] (d is the sensor diameter). Extra constrains are imposed to avoid overlap within the sensing region. We then design three possible groups of increasing conductivity values associated with the four frequencies as shown in Table~\ref{table:condGroup}, from which the changing conductivity values of target objects along frequency are assigned randomly. A distinct conductivity group is further ensured for each circular object within a phantom. This setup was adopted to simulate potential target application scenarios in tissue engineering (e.g. cell culture imaging \cite{dlgs:Chen}). 

A total of $4 \times 12,414$ (where 4 is the number of current frequencies) pairs of voltage-conductivity samples were generated through finite element modelling simulation. Considering phantom complexity, we generated $4 \times 3k$ one-object samples, $4 \times 4k$ two-object samples, $4 \times 5,414$ three-object samples. They are partitioned into $4 \times 8,700$ training set, $4 \times 1,900$ validation set, and $4 \times 1,814$ testing set for network training.

To eliminate the influence of systematic defect, we calibrate and normalize the voltage measurements and conductivity in the dataset, following:
\begin{equation} \label{eqvol}
\textbf{B}= \frac{\textbf{V}_{mea} - \textbf{V}_{ref}}{\textbf{V}_{ref}},
\end{equation}
\begin{equation} \label{eqcon}
\textbf{X}= \frac{{\pmb{\sigma}}_{mea} - {\pmb{\sigma}}_{ref}}{{\pmb{\sigma}}_{ref}},
\end{equation}
where ${\pmb{\sigma}}_{ref}$ and $\textbf{V}_{ref}$ denote the reference conductivity distributions and corresponding measurement data respectively with only background substance (discussed in Section \ref{mmvfoundation}); ${\pmb{\sigma}}_{mea}$ and $\textbf{V}_{mea}$ denote respectively the conductivity distribution and measurement with perturbations.

\begin{figure*}[!t]
\centerline{\includegraphics[width=\linewidth]{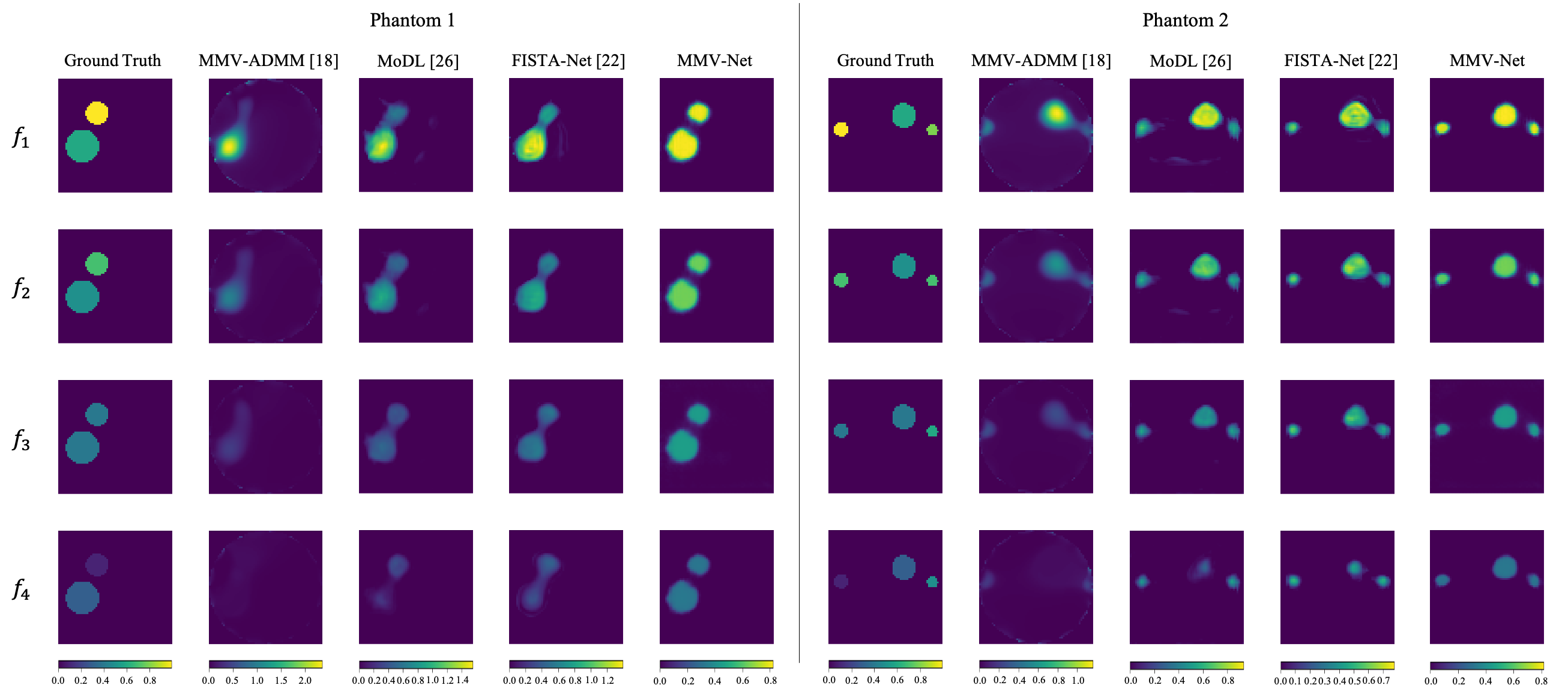}}
\caption{Comparison of the proposed MMV-Net with the state-of-the-art imaging approaches on two simulated phantoms from the testing set. }
\label{fig:phantomSta}
\end{figure*}

\subsection{Evaluation on Simulation Data}
In this sub-section, we evaluate the performance of the proposed MMV-Net using simulated mfEIT data.
\subsubsection{Performance Comparison}

\begin{table}[!t]
\caption{Performance comparisons (PSNR, SSIM and RMSE) on Edinburgh mfEIT Dataset.}
\begin{center}
\begin{tabular}{l  >{\centering}p{0.05\textwidth} >{\centering}p{0.06\textwidth} >{\centering}p{0.06\textwidth} >{\centering}p{0.06\textwidth} >{\centering\arraybackslash}p{0.06\textwidth}
}
\hline

\multirow{2}{*}{Metrics} & {\textit{Frequency}} & MMV- & \multirow{2}{*}{MoDL\cite{Modl:Aggarwal}} & {FISTA-} & \multirow{2}{*}{MMV-Net} \\
& {\textit{Channel}} & ADMM\cite{admmmmv:Zhang}& & Net\cite{FISTANet:Xiang}&  \\

\hline
\multirow{5}{*}{PSNR} & 1 & 19.3950 & 21.4738	 & 21.4444 & 	\textbf{23.7423} \\
& 2 & 22.0398 &24.1978 & 	24.4338 & 	\textbf{26.1284} \\
& 3 & 24.3093	&26.1114&	26.6960&	\textbf{28.5507} \\
& 4 & 26.6077	&27.0675&	27.6000&	\textbf{28.6125}\\
& \textit{Average} & 23.0880&	24.7126&	25.0435&	\textbf{26.7585} \\

\hline
\hline
\multirow{5}{*}{SSIM} & 1 & 0.4347&	0.8467&	0.8784&	\textbf{0.9354} \\
& 2 &  0.5266&	0.8676&	0.9092&	\textbf{0.9312}\\
& 3 &  0.6175&	0.8712&	0.9182&	\textbf{0.9469}\\
& 4 & 0.6383&	0.8569&	0.9092&	\textbf{0.9265}\\
& \textit{Average} &  0.5543&	0.8606&	0.9038&	\textbf{0.9350}\\

\hline
\hline
\multirow{5}{*}{RMSE} & 1 & 0.1175&	0.0910&	0.0933&	\textbf{0.0700} \\
& 2 & 0.0836&	0.0651&	0.0644&	\textbf{0.0527 }\\
& 3 & 0.0650&	0.0526&	0.0499&	\textbf{0.0404} \\
& 4 & 0.0549&	0.0512&	0.0496&	\textbf{0.0407}\\
& \textit{Average} & 0.0803&	0.0650&	0.0643&	\textbf{0.0510} \\

\hline

\multicolumn{2}{c}{No. of learning} & \multirow{2}{*}{NA} & \multirow{2}{*}{112,517} & \multirow{2}{*}{75,045} & \multirow{2}{*}{8,780} \\

\multicolumn{2}{c}{parameters} & & & &  \\

\hline
\multicolumn{6}{l}{{Best results are highlighted in bold.}}\\
\end{tabular}
\label{tab:comparisons}
\end{center}
\end{table}

we compare the proposed MMV-Net with three state-of-the-art image-reconstruction methods for mfEIT, i.e. MMV-ADMM \cite{admmmmv:Zhang}, MoDL \cite{Modl:Aggarwal}, and FISTA-Net \cite{FISTANet:Xiang} on the \textit{Edinburgh mfEIT Dataset}. MMV-ADMM is a conventional MMV-based method with the ADMM solver that can be adjusted and applied for mfEIT image reconstruction. MoDL and FISTA-Net are model based deep learning methods targeted at single measurement vectored based tomographic imaging. MoDL unrolls the traditional ADMM algorithm, while FISTA-Net is based on the FISTA framework \cite{fista:Beck}. The number of trainable parameters of the latter two model-based deep learning methods are given in the last row of Table~\ref{tab:comparisons}.

Table~\ref{tab:comparisons} shows quantitative comparisons based on average Peak Signal to Noise Ratio (PSNR), Structural Similarity Index Measure (SSIM), and Root Mean Square Error (RMSE) on all the testing data. MMV-Net outperforms all competing approaches at all frequencies. Note that there is an explicit improvement of PSNR, SSIM and RMSE from $f_1$ to $f_4$. This is due to the higher sensitivity of these metrics to larger conductivity contrasts.

Fig.~\ref{fig:phantomSta} demonstrates reconstructions of two simulated phantoms for qualitative comparison. MMV-ADMM can hardly reconstruct lower conductivity contrasts, especially at $f_3$ and $f_4$, whereas MoDL and FISTA-Net performs better. In contrast, the proposed MMV-Net can restore the most consistent structures/shapes and the conductivity changes along the frequency domain more smoothly. In addition, MMV-Net can distinguish fairly close objects more effectively than the other methods, which clearly demonstrates the advantages of SSA and ConvLSTM used in MMV-Net. However, all methods failed to yield accurate conductivity values of each object. Even the best performing MMV-Net tends to assign similar values to all objects, although the shapes estimated  are close to the ground truth. The potential reason is that the approximated linearization in \eqref{eq:mmv} is unable to handle such non-linear circumstances, i.e. the sensitivity matrix $\textbf{A}$ suffers from errors when interpreting conductivity levels from the measurement data.

\begin{table}[!t]
\caption{Ablation study on the validation set. (SSA: Spatial Self-attention Module; ConvLSTM: Convolutional LSTM Module.}
\begin{center}
\begin{tabular}{p{1cm} >{\centering}p{1.1cm} >{\centering}p{1.3cm}| l l l}
\hline

{Method}& SSA & ConvLSTM  & PSNR & SSIM & RMSE\\
\hline
\hline
\multirow{3}{*}{MMV-Net} &\checkmark & & 24.9132 &0.7640	&0.0622	\\

 &  &\checkmark &26.5144	&0.9300	&0.0530	\\

  &  \checkmark &\checkmark 	&{26.9817}	&{0.9364}&{0.0498}	\\
\hline

\end{tabular}
\label{table:Ablation}
\end{center}
\end{table}

\begin{figure}[!t]
\centerline{\includegraphics[width=.9\linewidth]{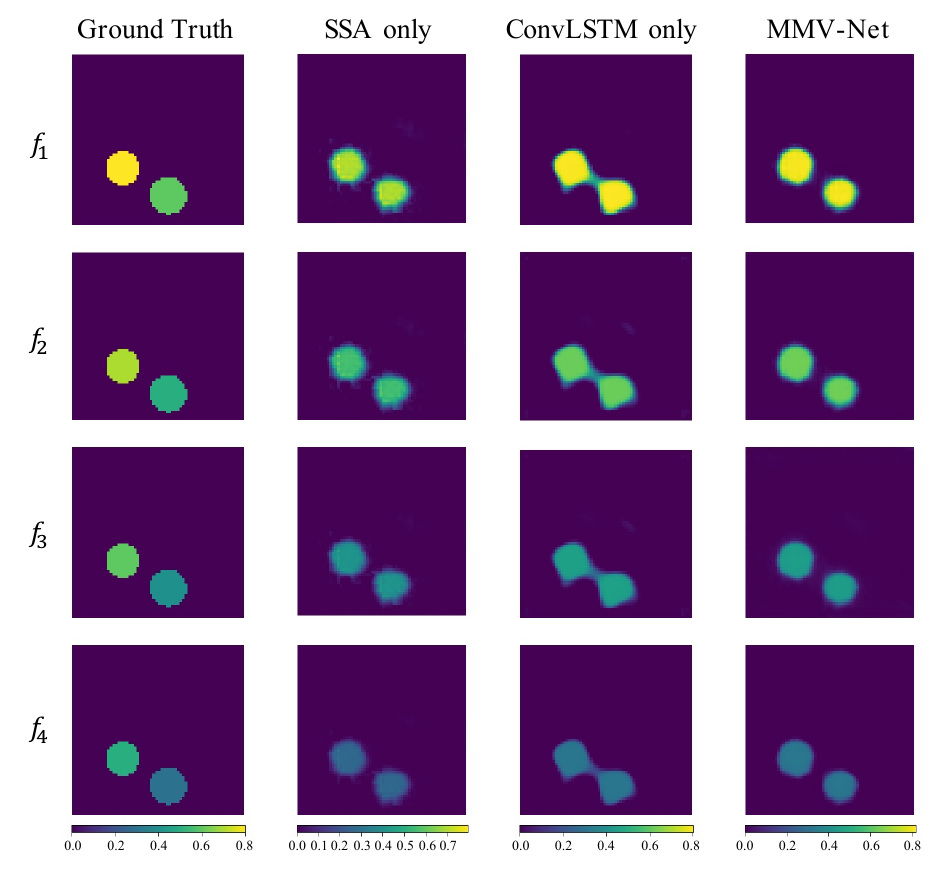}}
\caption{An example of image reconstruction results from ablation study. }
\label{fig:ablationstudy}
\end{figure}
\subsubsection{Ablation Studies}
In MMV-Net, we employ a Spatial Self-Attention (SSA) Module and a Convolutional LSTM (ConvLSTM) Module to capture intra- and inter-frequency correlations for high-performance mfEIT reconstructions. To verify the performance of the two modules, we conduct ablation studies (see Table~\ref{table:Ablation}). MMV-Net with only the ConvLSTM module outperforms MMV-Net with individually the SSA module by 6\% in PSNR, 22\% in SSIM, and 15\% in RMSE. Integration of the two modules into MMV-Net brings further improved performance of 26.9817 dB in PSNR, 0.9364 in SSIM, and 0.0498 in RMSE.

Fig.~\ref{fig:ablationstudy} illustrates the visual effects of the two modules. As expected, the SSA module itself manages to split the two objects and provide rough shapes but relatively vague boundaries, whereas the ConvLSTM module enhances the continuity but focuses less on shapes. Taking advantage of both modules, the proposed MMV-Net demonstrated superior performance among all.

\subsubsection{Iteration Analysis}

\begin{table}[!t]
\caption{Improvement in reconstruction quality on validation data with increasing number of iterations of the network.}
\begin{center}
\begin{tabular}{p{0.05\textwidth}  >{\centering}p{0.05\textwidth} >{\centering}p{0.05\textwidth} >{\centering}p{0.05\textwidth} >{\centering}p{0.05\textwidth} >{\centering\arraybackslash}p{0.05\textwidth}
}
\hline
{No. of } & \multirow{2}{*}{5} &\multirow{2}{*}{6} & \multirow{2}{*}{7} & \multirow{2}{*}{8} & \multirow{2}{*}{9}  \\
{Iterations}&  & & &  & \\
\hline
\hline
{PSNR} &26.8428	&26.9422&	26.9817&	27.0174&	27.0177\\

{SSIM} & 0.9341&	0.9342&	0.9364&	0.9391&	0.9394\\

{RMSE}  &0.0506&	0.0501&	0.0498&	0.0496&	0.0496\\
\hline

\end{tabular}
\label{table:IterationAnalysis}
\end{center}
\end{table}

\begin{figure}[!t]
\centerline{\includegraphics[width=.8\linewidth]{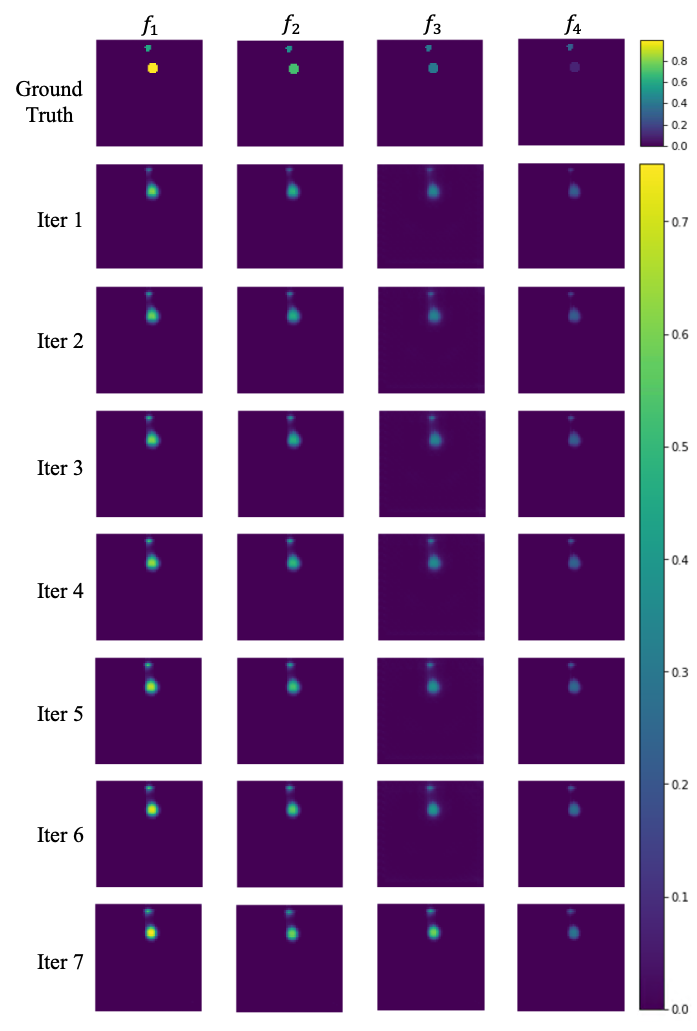}}
\caption{Intermediate reconstructed images by MMV-Net at different iterations. }
\label{fig:simuiter}
\end{figure}

Table~\ref{table:IterationAnalysis} shows the impact of the iteration number $ K_s$. It can be observed that average PSNR, SSIM, and RMSE values on the validation set are improved with increasing iterations. These improvements slow down considerably when $K_s\geq 7$. Therefore, we use $K_s = 7$ for configuration as a compromise of performance and computational cost.

We show the intermediate reconstructed images at all iterations in Fig.~\ref{fig:simuiter}. Each row corresponds to an iteration under different frequencies. The reconstruction quality improves gradually along iteration. More specifically, details in the structural information are clearer and more accurate while tiny changes in conductivity values raise as it goes deeper. It might be because we put more parameters in the SSA modules to learn the structural information.

\subsubsection{Generalization Ability}
\begin{figure}[!t]
\centerline{\includegraphics[width=.8\linewidth]{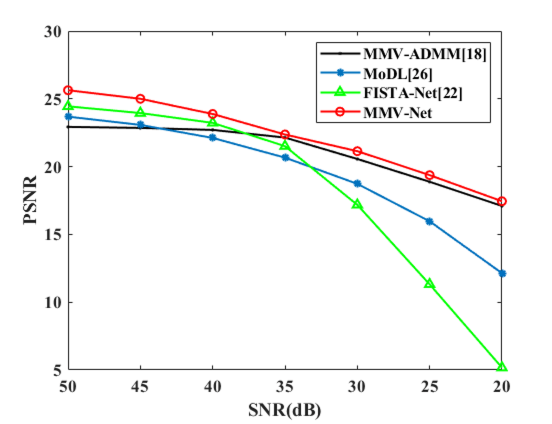}}
\caption{Generalization ability study of different noise levels.}
\label{fig:Generalization}
\end{figure}

We demonstrate the generalization ability of MMV-Net by adding different levels of noise to the measurement data and evaluate the image quality based on PSNR. Fig.~\ref{fig:Generalization} shows the average PSNR values of different methods. Degraded performance can be observed for all methods, whilst the proposed MMV-Net is the most robust against noise and FISTA-Net suffers a rapid decay. MoDL and FISTA-Net are more robust than MMV-ADMM with lower noise levels (e.g. 45dB) but MMV-ADMM exceeds both MoDL and FISTA-Net when the SNR is approximately smaller than 37dB. 

However, it is worth mentioning that all three learning-based methods are trained with only noise-free data. We believe our MMV-Net will retain more advantage upon MMV-ADMM and the other two model-based learning approaches should demonstrate more robustness to noise if sufficient noisy data are added in the training stage.

\subsubsection{Convergence Performance}

\begin{figure}[!t]
\centerline{\includegraphics[width=.9\linewidth]{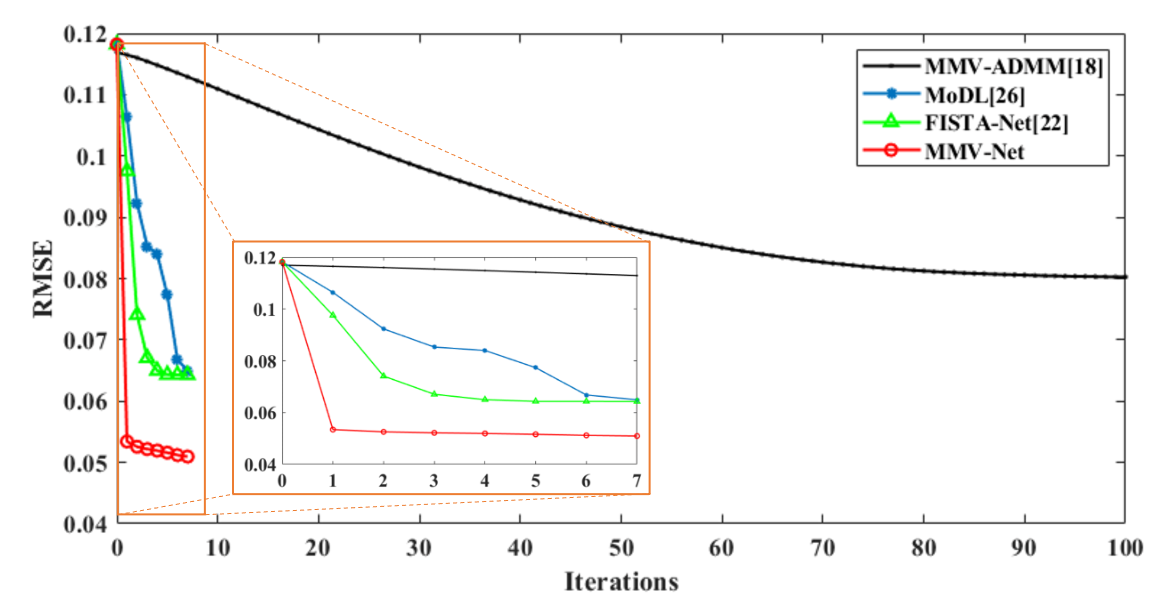}}
\caption{Convergence performance analysis on the testing set.}
\label{fig:Converge}
\end{figure}

Fig.~\ref{fig:Converge} illustrates the convergence performance of MMV-ADMM, MoDL, FISTA-Net, and MMV-Net. MMV-ADMM runs 100 iterations and ultimately converges to a certain level. While much faster convergence can be achieved by all network-based approaches. They adopts only 7 iterations and saturate at a much lower RMSE. It can also be observed that FISTA-Net performs smoother convergence than MoDL though they  finally stop at similar values. However, MMV-Net converges even faster than MoDL and FISTA-Net.

\subsection{Evaluation on Experimental Data}

\begin{figure*}[!t]
\centerline{\includegraphics[width=.85\linewidth]{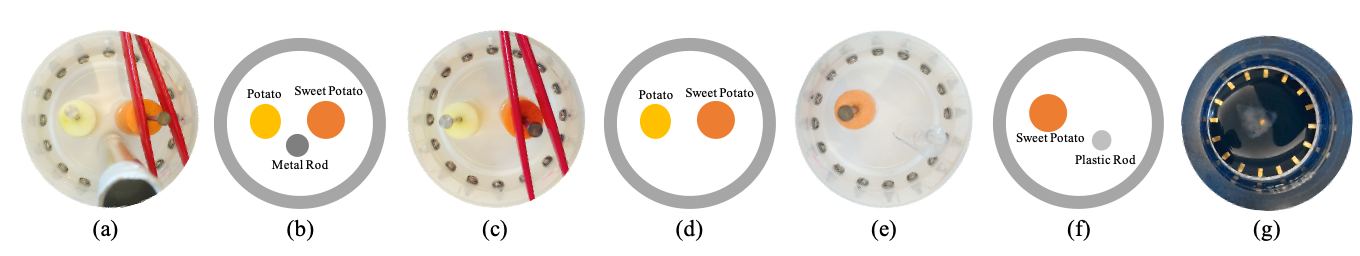}}
\caption{Experiment phantoms using two different 16-electrode EIT sensors. (a) Phantom 1: potato rod, sweet potato rod and metal rod. (b) Geometric distribution of phantom 1.
(c) Phantom 2: potato rod and sweet potato rod. (d) Geometric distribution of phantom 2. (e) Phantom 3: sweet potato rod and plastic rod. (f) Geometric distribution of phantom 3. (g) Phantom 4: MCF-7 cell pellet}
\label{fig:expphantom}
\end{figure*}

\begin{figure*}[!t]
\centerline{\includegraphics[width=.9\linewidth]{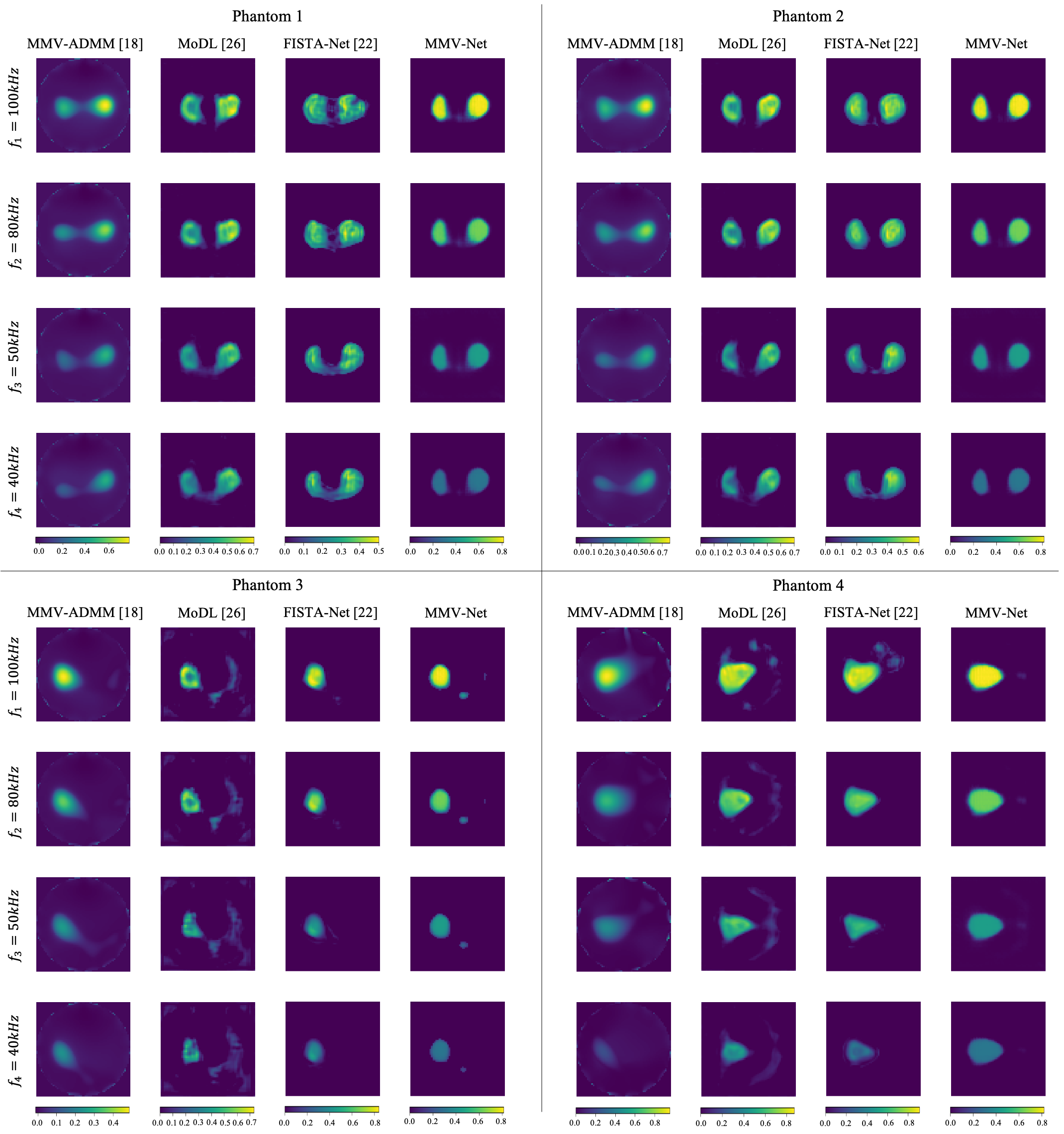}}
\caption{mfEIT image reconstruct results of four experimental phantoms. }
\label{fig:phantomComp}
\end{figure*}

In addition to simulation study, we carried out real-world experiments on two different EIT sensors \cite{mfeit:Yang,dlgs:Chen} to examine the generalization ability of the proposed method. The inner diameter of the first 16-electrode EIT sensor is 94 mm. We use a potato cylinder, a sweet potato cylinder, a metal cylinder and a plastic cylinder with different conductivity values as imaging targets. Fig.~\ref{fig:expphantom}(a)-(f) show pictures and corresponding geometric distributions of the three phantoms based on the first EIT sensor, which contain combinations of different targets. The background substance is saline with a conductivity of 0.07 $S\cdot m^{-1}$. The excitation frequencies are $ \{f_1, f_2, f_3, f_4, f_5\} = \{100, 80, 50, 40, 10\} kHz$, and $10 kHz$ is selected as the reference frequency. The conductivity of metal and plastic hardly changes with frequency, whilst the conductivity of potato and sweet potato increases progressively with the increase of current frequency \cite{mfeit:Yang}. The second miniature EIT sensor has 16 planar electrodes and an inner diameter of 15mm (see Fig.~\ref{fig:expphantom}(g)). The background substance is cell culture media with a conductivity of 2 $S\cdot m^{-1}$. The imaging object is a triangular MCF-7 human breast cancer cell pellet, which is less conductivity than the background substance and demonstrate an increasing conductivity with the increase of current frequency.

Fig.~\ref{fig:phantomComp} illustrates the mfEIT image reconstruction results based on experimental data. Overall, only MMV-ADMM and the proposed MMV-Net  manage to consistently provide a clear trend of conductivity values with respect to frequency. MMV-Net further produces more accurate shapes and less artifacts. Note that the potato cylinder and the sweet potato cylinder in experiment phantom 1 have the same location as that in experiment phantom 2, which is most successfully recovered by the MMV-Net. MMV, MoDL and FISTA-Net fail to identify the conductive metal cylinder in experiment phantom 1, whereas MMV-Net can roughly observe the metal cylinder but the shape is underestimated. For experiment phantom 3, MMV-ADMM, FISTA-Net and MMV-Net are more noise resistant than MoDL. However, MoDL and MMV-Net can reconstruct the non-conductive plastic cylinder. Similarly, for experiment phantom 4, MMV-Net is the most effective at inhibiting artifacts and shows more shape consistency at all frequencies. The results suggest that MMV-Net generalizes well to real-world experiments and outperforms the conventional model based method and state-of-the-art learning approaches by the competitive capability of capturing both intra- and inter-frequency correlations.

\section{Conclusion}\label{sec:Conclusion}
We proposed a model-based learning approach named MMV-Net to address the simultaneous image reconstruction problem of mfEIT. MMV-Net combines the advantages of the traditional MMV-ADMM algorithm and deep networks. All parameters are learned during training, rather than manually tuned.  We introduced the spatial self-attention module and convolutional LSTM module to learn both spatial and frequency correlations between mfEIT images. Ablation experiments showed that cascading both modules strengthened the structural information effectively and provided superior results. Simulation and real-world experiments demonstrated that the proposed MMV-Net outperformed the state-of-the-art methods in terms of image quality, generalization ability, noise robustness and convergence performance. This work can be readily extended to solve other tomographic image reconstruction problems. The 3D version of MMV-Net will also be investigated in the near future for 3D cell culture imaging.

\end{document}